\documentclass[aps,prc,nofootinbib]{revtex4}

\usepackage{graphicx}

\begin{document}
\begin{center}

{ \large\bf QCD-view on hadron form factors in space-like and time-like momentum transfer regions}

\vspace*{0.5 true cm}

Egle Tomasi-Gustafsson and  Michail P. Rekalo \footnote{ Permanent address:
\it NSC-Kharkov Physical Technical Institute, 61108 Kharkov, Ukraine}

\vspace*{0.5 true cm}
{\it DAPNIA/SPhN, CEA/Saclay, 91191 Gif-sur-Yvette Cedex,
France}
\end{center}
\vspace*{0.5 true cm}

\def\gms{$G_{Ms}~$}
\def\gmp{$G_{Mp}~$}
\def\gmn{$G_{Mn}~$}
\def\ges{$G_{Es}~$}
\def\gep{$G_{Ep}~$}
\def\gen{$G_{En}~$}

{\small 
\begin{center}
{\bf Abstract}
\end{center}

QCD gives definite predictions for hadron electromagnetic form factors in space-like and time-like momentum transfer regions,  such as the quark counting rules, the hypothesis of hadron helicity conservation, and the relations between nucleon and deuteron form factors in the formalism of reduced nuclear matrix elements. Recent precise data about these form factors, obtained in polarization experiments at the Jefferson Laboratory, have essentially changed our view on this subject. QCD-predictions do not apply to these data up to $Q^2$=5-6 GeV$^2$ for deuteron and for the electric form factor of proton. An analysis of these data suggests that the asymptotic region will more probably start at $Q^2$=20-25 GeV$^2$. We show that the separation of magnetic and electric proton form factors in the time-like region represents the most stringent test of the asymptotic regime and QCD-predictions.
}
\vspace*{0.5 true cm}

In this talk we will discuss the recent developments in the field of hadron electromagnetic form factors (FFs), due to the very precise and surprising data obtained at the Jefferson Laboratory (JLab), in $\vec e+p\to e+\vec p$ elastic scattering.

The application of the polarization transfer method, proposed about 30 years ago \cite{Re68} has been possible only recently, as it needs high intensity polarized beams, large solid angle spectrometers and advanced techniques of polarimetry \cite{Mi98,Jo00,Ga02} in the GeV range.

At these energies, one probes distances of the order of the nucleon size or less, and  would expect expect the manifestation of quark degrees of freedom.

FFs, which characterize the internal structure of composite particles, can be experimentally measured and theoretically calculated, thus providing a good playground for the models of nucleon structure.

In a P and T invariant theory, a particle with spin $S$ is characterized by $2S+1$ electromagnetic form factors, which are complex (real) functions in the time(space)-like region. The nucleon has two FFs, called electric $G_{EN}$ and magnetic $G_{MN}$, which, a priori, are different.

Before the new, precise data appeared in the space-like (SL) region, the dipole approximation, $G_D=\left [1+Q^2/a \right ]^{-2}$ with $a=0.71 ~\mbox{GeV}^2$, has been considered a good approximation for three of the four nucleon FFs,  \gen being neglected or taken according to \cite{Galster}. 

In the Breit system, FFs are related to the Fourier transform of the charge and magnetic moment
distributions and the dipole approximation results from an
exponential distribution, the coefficient $a=0.71$  GeV$^2$ corresponding to a root mean squared radius $\sqrt{<r^2>}$=0.81 fm.

Perturbative QCD gives definite rules about the
scaling behavior of the form factors and about helicity conservation, indicating where quark degrees of freedom should be taken explicitely into account. Scaling laws, which give the probability that a hadron remains intact after absorbing a photon of momentum $Q^2$, have been formulated in \cite{Br76} as $F=\left [1+Q^2/(n\beta_n^2 )\right ]^{n-1}$, where $n$ is the number of quarks and $\beta_n$ the quark average momentum. When applied to proton, they give the same power law as dipole, including the coefficient $a=0.71$, derived from a fit to pion form factors.

The traditional way to access FFs in the SL region, is the elastic scattering of electrons on hadrons $e+h\to e+h$, and in the time-like (TL) region, the annihilation processes $e^++e^-\leftrightarrow p+\overline{p}$. The unpolarized elastic $ep$ cross section, in one-photon exchange approximation, can be  written as a function of the electric \gep  and magnetic \gmp proton form factors \cite{Ro50}:
\begin{equation}
\displaystyle\frac{d\sigma}{d\Omega}(ep\to ep)=
\left (\displaystyle\frac{d\sigma}{d\Omega}\right)_{Mott}\left [ \displaystyle\frac {G_{Ep}^2+\tau G_{Mp}^2}{1+\tau} +2\tau
G_{Mp}^2\tan^2\displaystyle\frac{\theta_e}{2}\right ],~ \tau=\displaystyle\frac{Q^2}{4M^2} 
\label{eq:ros}
\end{equation}
with
$$
\left (\displaystyle\frac{d\sigma}{d\Omega}\right)_{Mott}=
\frac{\alpha^2~\cos^2(\theta_e/2)E'}{4E^3\sin^4(\theta_e/2)} \mbox{~and~}
E'=\displaystyle\frac {E}{ 1+2\displaystyle\frac {E}{M}\sin^2(\theta_e/2)}$$
where $M$ is the proton mass, $E$ is the energy of the incident electron,
$E'$ and $\theta_e$ are the energy and scattering angle of the outgoing electron,
$\alpha$ is the fine structure constant. The momentum of the virtual photon is $Q^2=4EE'\sin^2(\theta_e/2)$ and it is positive in the SL region.

In the TL region the cross section can be expressed as a function of
FFs according to the following formula \cite{Zi62}:
\begin{equation}
\displaystyle\frac{d\sigma}{d(cos\theta)}(e^+e^-\to p\bar p) =
\displaystyle\frac{\pi\alpha^2}{8M^2\tau\sqrt{\tau(\tau-1)}}
\left [ \tau |G_{Mp}|^2(1+\cos^2\theta)+|G_{Ep} |^2\sin^2\theta
\right ],
\label{eq:tl}
\end{equation}
where $\theta$ is the angle between the electron and the antiproton
in the center of
mass frame.

Eqs. (\ref{eq:ros}) and (\ref{eq:tl}), contain the moduli squared of the FFs, therefore one can not access their sign. Moreover the contribution of \gmp  appears weighted by the factor $\tau$: as $Q^2$ increases, it becomes the dominant term, making the extraction of \gep  very imprecise. 

In the SL region the collisions of polarized electron and polarized target (or measuring the polarization of the scattered proton), induce an interference term proportional to the product $G_{Ep}G_{Mp}$. It is therefore more sensitive to a small contribution of \gep . 

Data have been obtained at JLab up to $Q^2=5.6$ GeV$^2$ and an extension up to 9 GeV$^2$ is in preparation \cite{00111}. The  measurement of the polarization of protons at momentum up to $P_p=5.3$ GeV/c can be done with a POMME-like polarimeter, as it has been proved at the JINR-LHE, in Dubna \cite{dubnacalib}. Although the analyzing power for the inclusive reaction $p+CH_2\to \mbox{one~ charged~particle ~}+~X$  decreases with increasing incident momentum, it is still sizeable at a proton momentum of 5.3 GeV.

The recent data \cite{Jo00,Ga02} show a linear deviation from the dipole behavior and can be parametrized by $ \mu $ \gep  / \gmp $=1.0-0.130(Q^2/\mbox{GeV}^2-0.04)$ \cite{Ga02} ($\mu$ is the proton magnetic moment), definitely proving that the electric and magnetic distributions in the proton are different.

\begin{figure}
\begin{center}
\includegraphics[width=12cm]{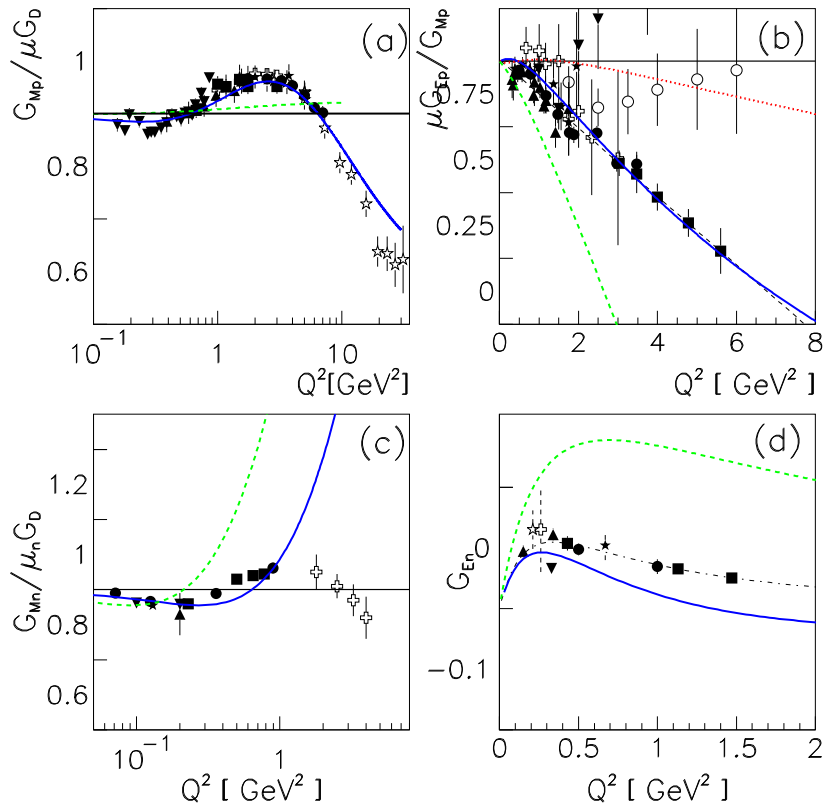}
\caption{\label{fig:fig1} Nucleon FFs in the SL region. The magnetic FF for proton (a) and for neutron (c) are normalized to one and divided by $G_D$. The proton electric FF is shown as the ratio $\mu $\gep /\gmp (b), the dotted line is from \protect\cite{Wataghin}. The neutron electric FF  (c) is plotted with the \protect\cite{Galster} parametrization (dash-dotted line). The solid line is from \protect\cite{IJL}, the dashed line from \protect\cite{Zi66}.  }
\end{center}
\end{figure}

Although different models can reproduce this trend, (as constituent quark models \cite{Ca00}, soliton model \cite{Ho96}, diquark model \cite{Kroll} ..)  few of them give a satisfactory description of all four electromagnetic form factors and even fewer of them are applicable in SL and TL regions. We would like to quote here the models based on vector meson dominance (VMD)\cite{GK92}. The most recent one \cite{Lomon}  contains several parameters to fit the world data and  includes the asymptotic QCD behavior.  Note that 'Il Nuovo Cimento' reported  the first studies, in this field. In Ref. \cite{Wataghin} the right trend for the \gep  was already proposed, and in \cite{Zi66} a  'one parameter fit'  gave a good qualitative description of the present data and prescriptions for TL  region as well. The best VMD-predictions, concerning the proton FFs,  are given by \cite{IJL}  (Fig. \ref{fig:fig1}). A compilation of the world data reported in the figure can be found in \protect\cite{Lomon}.

The nucleon FFs are important ingredients for the calculations of the light nuclei structure. One of the consequences of the new data is the revision of the models of the deuteron structure. 
Following \cite{Br76}, let us introduce a generalized deuteron FF, $F_D(Q^2)$, $F_D(Q^2)=\sqrt{A(Q^2)}$, where $A(Q^2)$ is the structure function related to the forward electron deuteron cross section, and a reduced deuteron FF $f_D(Q^2)$:
\begin{equation}
f_D(Q^2)=\displaystyle\frac{F_D(Q^2)}{F_N^2(Q^2/4)},
\label{eq:eq1}
\end{equation}
where $F_N$ is the nucleon electromagnetic FF. The $Q^2$-behavior of
$f_D(Q^2)$ (at large $Q^2$) can be predicted in the framework of pQCD, in the following form:
\begin{equation}
f_D(Q^2)=N\displaystyle\frac{\alpha_s(Q^2)}{Q^2}\left ( ln \displaystyle\frac{Q^2}{\Lambda^2}\right)^{-\Gamma},
\label{eq:eq2}
\end{equation}
where N is the normalization factor (which is not predicted by QCD), $\alpha_s$ is the running QCD strong interaction coupling constant, $\Lambda$ is the scale QCD parameter, and $\Gamma$ is determined by the leading anomalous dimension, here $\Gamma=-\displaystyle\frac{8}{145}$.

In \cite{Al99} it was shown that the QCD prediction (\ref{eq:eq2}), which can be applied to asymptotic momentum transfer, is working well already for $Q^2\ge 2$ GeV$^2$, with a plausible value of the parameter
$\Lambda \simeq$ 100 MeV, in agreement with the values determined by many other possible methods \cite{pdg}.

In  \cite{Br76} another interesting prediction, concerning the scaling behavior of the reduced deuteron FF, was done:
\begin{equation}
f_R=\left ( 1+\displaystyle\frac{Q^2}{m_0^2}\right )f_D(Q^2)\simeq const,
\label{eq:eq3}
\end{equation}
where $m_0^2=0.28$ GeV$^2$ is a parameter related to the $Q^2$-behavior of the  pion FF. The same data from \cite{Al99}, if plotted in the representation of the reduced deuteron FFs, should illustrate the $Q^2$-independence of this product.

\begin{figure}
\begin{center}
\includegraphics[width=8cm]{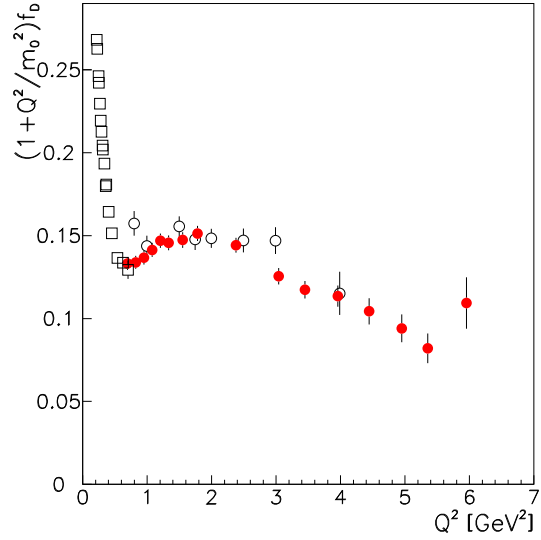}
\caption{\label{fig:fig2} Data set corresponding to the reduced deuteron FFs multiplied by $(1+Q^2/m_0^2)$. Open circles are from \protect\cite{Ar86}, open squares from \protect\cite{Pl90}, solid circles from \protect\cite{Al99}.}
\end{center}
\end{figure}

This result was confirmed by the previous $A(Q^2)$ data \cite{Ar86}, in the limit of their accuracy. In Fig. (\ref{fig:fig2}) we  show  that the new, more precise data about $A(Q^2)$ \cite{Al99}, are not consistent with the prediction (\ref{eq:eq3}) as they show an evident dependence of the product $f_R$ on $Q^2$, even for $Q^2\ge $2 GeV$^2$. This behavior can not be changed by varying the parameter $m_0$. 

Although the scaling laws seem to be consistent with cross section measurements, up to 6 GeV$^2$, if one replaces the dipole approximation with other descriptions of the nucleon FFs, taking into account the deviation from dipole for \gep, a fit following Eqs. (\ref{eq:eq1}) and (\ref{eq:eq2}) shows a large instability for $\Lambda$ \cite{MPR03}.

One should mention that a calculation based on the impulse approximation, where the nucleon FF are taken from \cite{GK92}, satisfactorily  reproduces the existing data on the three deuteron FFS \cite{ETG01a}.

We will go a step further, and look for a definition of the asymptotic region with respect to the analyticity properties of complex functions. 

FFs must obey the  Phr\`agmen-Lindel\"of
theorem \cite{Ti39}, which  gives a rigorous prescription for the asymptotic behavior of
analytical functions:
$\lim_{Q^2\to -\infty} F^{(SL)}(Q^2) =\lim_{Q^2\to \infty}
F^{(TL)}(Q^2)$.
This means that, asymptotically, the TL phase vanishes  and the real part of the FFs,
${\cal R}e  F^{(TL)}(Q^2)$, coincides with the
corresponding value $F^{(SL)}(Q^2)$.

\begin{figure}
\begin{center}
\includegraphics[width=10cm]{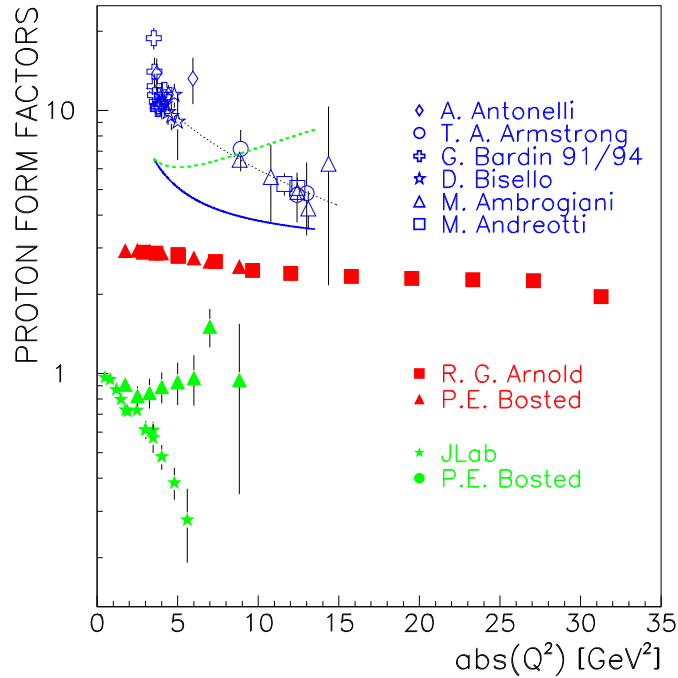}
\caption{\label{fig:aztl} Data for electric and magnetic FFs in
SL (solid symbols) and TL (open symbols) regions, scaled by dipole,
as functions of the modulus of $Q^2$. The dotted line is a fit to the TL data, according to
the functional form described in the text.The solid (dashed) line is the prediction for \gmp (\gep ) from \protect\cite{Zi66}.}
\end{center}
\end{figure}

The Rosenbluth
separation has not yet been realized in the TL region. In order to extract the FFs,  due to the poor statistics, it
is necessary to integrate the differential cross section over a wide angular
range. One assumes that the \gep -contribution plays a minor role in the cross
section and the
experimental results are usually given
in terms of $|G_{Mp}|$, under the hypothesis that \gep =0  or $|G_{Ep} |=|G_{Mp} |$. The first hypothesis is arbitrary. The second hypothesis is strictly
valid at threshold only, but there is no
theoretical argument which justifies its validity at any other momentum
transfer, where $s\neq 4M^2$. The prediction for the TL region in  \cite{Zi62}  shows that if \gep/$G_D$  decreases, \gmp/$G_D$   increases, due to the definitions (dashed and solid line, respectively in Fig. \ref{fig:aztl}).

Therefore a comparison of data in TL and SL region should give an unambiguous indication on the asymptotic region. 
The experimental data are shown in Fig. \ref{fig:aztl}, normalized to the function $G_D$. For a compilation see \protect\cite{ETG01}, here updated with recent data \cite{An03} (open squares).

The values of \gmp in the TL region,
obtained under the assumption  that $|G_{Ep} |=|G_{Mp} |$ (open symbols), are larger
than the
corresponding SL values (solid squares and solid triangles). This has been
considered
as a proof of the non applicability of the Phr\`agmen-Lindel\"of theorem,
or as an evidence that  the asymptotic regime is not reached \cite{Bi93}.

The magnetic form factor of the proton in the TL region (which is deduced
from the hypothesis \gep =0 (case 1) or \gep =\gmp (case 2), can be parametrized as:
$G_{M}^{(TL)}=G_D a/\left
(1+{s}/m_{nd}^2\right)$, where $a$ is a normalization
parameter and $m_{nd}^2=3.6\pm 0.9$ GeV$^2$ characterizes the deviation
from the usual dipole $s$-dependence. The extrapolation to higher $s$ based on
this formula (Fig. \ref{fig:aztl}, dotted line), indicates that the
Phr\`agmen-Lindel\"of theorem
will be satisfied by this FF, only for $s(Q^2)\ge 20$ GeV$^2$.

\begin{figure}
\begin{center}
\includegraphics[width=10cm]{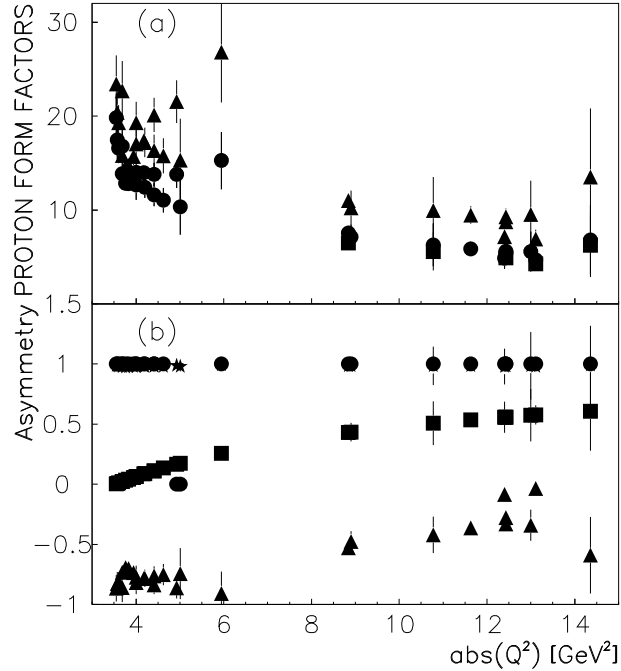}
\caption{\label{fig:asym} Nucleon form factors (a) and angular asymmetry (b) in TL 
region, 
deduced from the data according to different assumptions (see text).}
\end{center}
\end{figure}

Let us assume now that one of the 
two proton electromagnetic FFs  has reached the asymptotic regime and apply the Phr\`agmen-Lindel\"of theorem
to extract the other. This looks as  a reasonable hypothesis for $G_M$, which shows an early scaling behavior, in accordance with quark counting rules.
From Eq. \ref{eq:tl} we can 
deduce $|G_E|$, using  the existing 
experimental data about $\overline{p}+p\leftrightarrow e^+ +e^-$ (case 3). We report, in  Fig. \ref{fig:asym}, the recent  data in TL region, reanalized 
following the  possibilities suggested above. Fig.  \ref{fig:asym}a shows the values 
of the form factors taking $G_E=0$ (circles) and $|G_E|=|G_M|$ (squares) respectively). For case 3, where $G_M=G_D$, the values of 
$|G_E|$ (triangles) are larger than in cases 1 and 2. This suggests that asymptotics are not reached for $G_E$, as the values in the 
SL and TL regions get more apart. A 
fourth possibility is taking for 
$G_E$ in the TL region the values from \cite{Jo00,Ga02} and calculate 
$|G_M|$ (case 4). This affects very little the values of $G_M$, due to 
the kinematical factor 
$\tau$, which weights the magnetic contribution to the differential cross 
section (stars).

One can express the angular dependence of the differential cross section as a 
function of the angular asymmetry ${\cal A}$ as:
\begin{equation}
\displaystyle\frac{d\sigma}{d(\cos\theta)}=
\sigma_0\left [ 1+{\cal A} \cos^2\theta \right ], \mbox{~with~} {\cal A}=\displaystyle\frac{\tau|G_M|^2-|G_E|^2}{\tau|G_M|^2+|G_E|^2}, 
\end{equation}
where $\sigma_0$ is the value of the differential cross section at 
$\theta=\pi/2$. Fig. \ref{fig:asym}b shows the angular asymmetry for the different cases. Case 1 and case 2  
give, 
respectively,  ${\cal A}=1$ and ${\cal  A}=(\tau-1)/(\tau+1)$. The calculated 
asymmetries are very sensitive to the different underlying 
assumptions, therefore a precise measurement of this quantity would be very 
interesting. 

Finally, we note that the angular dependence of the cross section, Eq. (\ref{eq:tl}), results 
directly from the assumption of one-photon exchange, where the spin of the 
photon 
is equal 1 and the electromagnetic hadron interaction satisfies the 
$C-$invariance. 
Therefore the measurement of the differential 
cross section at three angles (or more) would also allow to test the presence of $2\gamma$ exchange. The relative role of the $2\gamma$ mechanism can increase at relatively 
large momentum transfer in SL and TL  regions, for the same physical reasons, 
which 
are related to the steep decreasing of the hadronic electromagnetic FFs, as 
recently discussed in \cite{Re99}.

\hspace*{0.2 truecm}
\begin{center}

{ \bf ...instead of Conclusions}
\hspace*{0.2 truecm}
\end{center}

{\it 
It is for me a great pleasure and honour to express my gratitude to Prof. R. A. Ricci, to whom this Conference is dedicated. As supervisor of my 'Tesi di Laurea, Universit\`a di Padova,  Anno Accademico DCCLVI', he gave me the basis to progress in the field of Nuclear Physics. Since that time, his enthusiasm, his dynamic personality, his optimism in affording a variety of problems of different nature, his broad interests and the ability to listen, to understand and synthesize information very rapidly, have always been for me a source of motivation and example. I take this opportunity to thank Prof. Ricci, and to wish to him, besides the recognizements due by the italian and world-wide nuclear physics community, to fully enjoy a time (may be) more quiet and to have a lot of satisfaction in his family life.
\begin{flushright}
E. T.-G.
\end{flushright}
}

{}

\end{document}